# Demonstration of Design verification model of Rubidium Frequency Standard

Bikash Ghosal, Sarathi Mandal, S.S Raghuwanshi**, Savita Singh**, A.Banik, R.K. Bahl,
K.S.Dasgupta,* G.M.Saxena**

*Space Applications Center, Indian Space Research Organization*
*Ahmedabad, Gujarat, INDIA*

*IIST, Trivendrum India
**NPL, New Delhi India (g m saxena e-mail: gmsnpl@yahoo.co.in)

**Abstract**
*In this paper we report the development of the design verification model (DVM) of Rb atomic frequency standard for the Indian Regional Navigational Satellite System(IRNSS) programme. Rb atomic clock is preferred for the space applications as it is light weight and small in size with excellent frequency stability for the short and medium term. It has been used in all other similar navigation satellite systems including GPS,GLONASS Galileo etc. The Rb atomic frequency standard or clock has two distinct parts. One is the Physics Package where the hyperfine transitions produce the clock signal in the integrated filter cell or separate filter cell configuration and the other is the electronic circuits which include frequency synthesizer for generating the resonant microwave hyperfine frequency, phase modulator and phase sensitive detector. In this paper the details of the Rb Physics package and the electronic circuits are given. The reasons for the mode change in Rb lamp have been revisited. The effect of putting the photo detector inside the microwave cavity is studied and reported with its effect on the resonance signal profile. The Rb clock frequency stability measurements have also been discussed.*

## I. INTRODUCTION

A rubidium frequency standard is a very versatile atomic clock with excellent short term frequency stability and it is relatively small in dimensions and light in weight. It is a passive atomic clock i.e., in this atomic clock a VCXO frequency is locked to the ground state hyperfine transition frequency of $Rb^{87}$ atoms. The atomic resonance serves as a discriminator and produces an error signal that varies in magnitude and phase as a function of the difference in frequencies between the applied RF excitation derived from the VCXO and the atomic hyperfine transitions. This error voltage is obtained and processed by a servo amplifier, which generates a voltage that controls the frequency of the crystal oscillator. The functional block diagram, detailed view of the Physics Package and tested DVM for the standard are shown in fig.1 (a), (b) and (c) respectively. The present work on the Rb atomic clock is related to Indian programme on the satellite Navigation satellite system-IRNSS. Under the programme, initially two Rb Physics packages have been developed and tested with in-house developed electronics. The simulation work on the thermal and vibration tests have also been completed as the final goal is to develop space qualified Rb atomic clocks. IRNSS programme is very similar to the navigation satellite systems of other countries.However,in the first phase of the programme only Rb atomic clocks will be used to achieve a accuracy of 20m over India and 1500km around it.

## II Physics Package

The core of the physics package consists of the rubidium bulb, integrated filter cell and the microwave cavity. The physics package also includes a solenoid providing necessary C-field, a triple magnetic shield layers to reduce effect of environment magnetic field fluctuations and two bifilar heaters to control the temperature of the physics package at the level of $0.1^0C$ or better. The optical resonance radiation from a rubidium lamp is transmitted through a glass cell containing $Rb^{87}$ isotope in vapour state and a buffer gas at low pressure, purpose of the latter being to reduce Doppler broadening in the microwave resonance of interest [1].The action of the optical resonance radiation produces a population inversion between the hyperfine levels in the ground state of $Rb^{87}$ far in excess of that which is obtained under thermal equilibrium. This process is called optical pumping [2]-[3].

The Rb lamp consists of electrode-less Rb bulb and a lamp exciter. The Rb bulb is made of Pyrex 7070 or 7740 glasses in spherical shape of outer diameter 10mm approx., wall thickness 0.5 to 0.7mm. The bulb is filled with Rb87 isotope and natural Rb of 99.99% purity in 1:1 ratio and each in the quantity of 0.350 mg (aprox). The use of the mixture of natural and $Rb^{87}$ in equal proportions ensures minimization of the light shift. The bulb is filled with Krypton gas at 2.0 ± 0.2 Torr of nearly 99.995% purity for ease of excitation as Krypton has low ionization potential. The bulb is excited by 55 - 100 MHz/2-3 watt RF exciter which is a Colpitt's oscillator run on a D.C power supply of rating 20Volt to 24volt and 0.2 to 0.4 Amp current. The light intensity and mode of operation of the lamp may be controlled by changing the gain and frequency of the oscillator with the help of a resistor and capacitor respectively. As the Rb lamp is highly critical component of the Physics Package it is necessary to keep provision to control these parameters precisely.

The lamp is self r-f heated but it requires additional temperature controller for stabilization of its intensity and the operation of the lamp in the proper mode. When the temperature is around $110^0$C the lamp operates in ring-mode [4, 5] i.e., the Rb light appears to be emitted from a narrow ring close to the surface of the bulb and the emitted purple colored Rb light has very narrow line width and there is no self-reversal [6]. The self-reversal is a phenomenon in which the spectral line shows a depression in the center of the line profile. When the temperature is increased, the intensity decreases and the color of the light becomes deep purple or reddish. The spectral lines become broad and highly self reversed. In view of the above-mentioned behavior of the lamp at different temperatures, it is desirable that the lamp's temperature should be stabilized at $110^0$C with temperature stability better than 0.1 Celsius. We shall discuss the reasons of the transition from the ring to the red mode. The phenomenon has been studied by various groups. Shah R.S [5], and Camparo [4] have described the mode transition due to the radiation trapping. Their explanations are not fully satisfactory.Camparo has made ad hoc assumption that P-levels of Rb atoms become metastable. And it gives rise to radiation trapping. The mode transition is revisited in this paper

We explain that to describe the Rb lamp mode transition, we should consider the coherent population trapping (CPT) [3] of Rb atoms in the ground state hyperfine levels. To elaborate on it, we consider Dicke's [1] seminal work on the presence of coherence in the spontaneous emission process. The coherence in the spontaneous emission leads to confinement of unexcited Rb atoms to the ground state (g.s) hyperfine sub-levels due to the destructive interference between the two radiation lines originating out of the transitions from ground state hyperfine levels $^5S$ F=2 and F=1 to $^5P_{1/2}$ in the Rb lamp. This trapping of some of the atoms in the g.s hyperfine levels is equivalent to the situation as if there is radiation trapping in the excited state, as inferred by Camparo. However, Camparo has made a assumption that the excited state behaves as metastable state. We are of the opinion, as mentioned above, that this reduced intensity of the Rb lamp may be the result of the trapping of the atoms in the ground state hyperfine levels due to the partial coherence in the spontaneous emission by the Rb lamp.Besides,in the integrated filter cell technique, the filtering of the undesired light may not be that efficient. So the presence of the coherent part in the undesired radiation line as well as in the desired radiation line may coherently trap some of the $Rb^{87}$ atoms in the ground state. Besides, the trapping of the atoms due to CPT is physically manifested in the loss of optical pumping efficiency in the Rb absorption cell, used in the Rb atomic clocks. On the basis of the above analysis, it is established qualitatively that the radiation trapping phenomenon is basically produced due to the CPT of the atoms both in the Rb bulb and absorption cell. The quantitative analysis of the radiation trapping due to CPT requires separate detailed experimentation and it is beyond the scope of this paper. However, the above discussions are quite elaborate and qualitatively sufficient to explain that radiation trapping results due to CPT of the atoms.

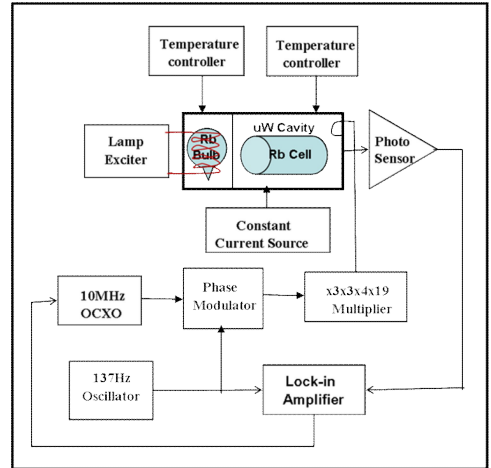

Fig.1 (a) - Functional block diagram of rubidium atomic frequency Standard

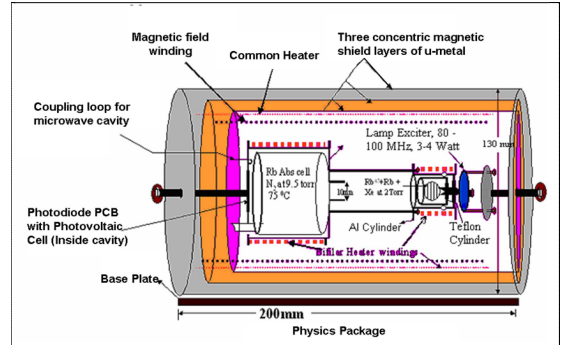

Fig.1 (b)-Detailed view of the Physics Package

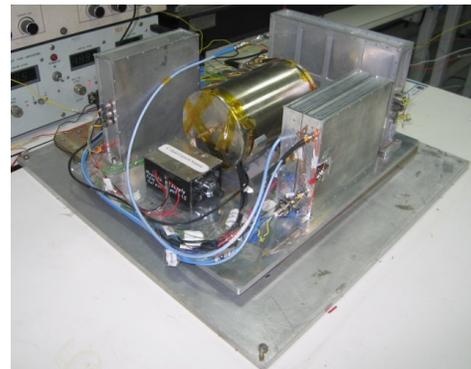

Fig.1 (c) - Tested DVM of rubidium atomic frequency standard

In the Rb Physics Package the residual temperature coefficient generally is of the order of $10^{-10}$ $^0C^{-1}$. Therefore, for achieving the frequency stability of the order of $10^{-13}$ or better the lamp and cell temperatures should have stability of the order of the fractions of Celsius. To control the temperature of the Rb lamp and the integrated Rb absorption cell separate heaters are provided. These heaters with proportional current controls have bifilar windings for eliminating the generation of the magnetic field by the current passing through the heaters. The glass-encapsulated thermistor is placed beneath these heaters for sensing the temperature and providing the necessary input to the temperature controller circuits. In addition to these two heaters, one common heater for maintaining the temperature a few degrees above the ambient temperature is provided. This double oven approach ensures temperature stability of a small fraction of a Celsius. The heaters are powered by 15 and 20Volt D.C supplies.

The integrated filter cell contains the natural rubidium and Nitrogen as a buffer gas. The cell is made of Pyrex glass so that the quality factor of the cavity remains practically unaffected. The Nitrogen gas also provides quenching of the scattered and fluorescence radiation. This property helps in preventing radiation trapping which tends to affect the efficient optical pumping of the Rb atoms [7].

An aluminum alloy (Al 6061) $TE_{111}$ microwave cavity tuned at 6.83468...GHz is used for resonant transitions of Rb atoms between the hyperfine levels in the ground state. Its dimensions are L= 40.00 mm (including tuning length of 10mm), inner diameter = 27.00 mm and wall thickness =7mm. The microwave feed is through a loop of 1.00mm dia. The absorption cell and the array of the photovoltaic cells are also kept inside the microwave cavity. The ground plane of the array of photovoltaic cells acts the termination of the microwave cavity. On the other end of the cavity is a tuning plunger with suitable threads for tuning the cavity precisely. The microwave signal has its magnetic field in the direction of the optical axis. This is a requirement for exciting the clock transitions between F=2, $m_F$=0 — F=1, $m_F$=0 which have second order dependence on the magnetic field. The loaded Q of the microwave cavity is more than 500.The cavity is provided with bifilar heater winding for maintaining the temperature of the microwave cavity and absorption cell at $75^0C$ with temperature stability of better than $0.1^0C$. The absorption cell is kept at a temperature of $75^0C$ for obtaining nearly zero light shifts.

**Effect of Placing Photodiode PCB inside microwave cavity:**

To detect the hyperfine transitions and to obtain the clock signal, a number of photovoltaic cells, with high efficiency in the near infrared (800nm) region are mounted on a PCB. The back side of the PCB on which these photodiode are mounted is a ground plane. The simulation results with the photodiode PCB inside the microwave have shown that the ground plane acts as microwave cavity termination and the length of the microwave cavity is negligibly increased. But the microwave leakage is reduced as no hole in the microwave cavity on the detector side for collecting transmitted light is required. This may result in slight increase in cavity Q. The photo detectors cover the cross-sectional area of the absorption cell. The detected light signal is of the order of 130-170 milliVolt at $75^OCelsius$ and at the resonance, the dip in the light intensity of 150-200 µVolt has been observed at the output of the photo detector.

A small constant D.C magnetic field is applied to the Rb atoms in the absorption cell in order to provide a quantum axis along the Rb light for the field independent F, m=0 - F' m'=0 transitions. It also provides some leverage for manipulating the hyperfine energy levels so that the hyperfine transition frequency could be exactly matched with the applied microwave frequency at the resonance. Besides, some uncertainty in the buffer gas pressure may also be taken care of by the magnetic field. The constant D.C magnetic field is produced by a solenoid. The winding is done on a Hylem cylinder, a non magnetic material. Its diameter is 70mm and length is 120mm.The winding is compact so that the homogeneity of the magnetic field is maintained .A field of 250-450miligause is produced by the solenoid.A precision constant current source has been made to control the current in the solenoid.

The main parameters of physics package are given in table 1.

Table 1
Lists the main parameters of the physics package

| Cavity dimensions | O.D=54mm,L=56mm,ID=32 mm,tuning L=30-35mm |
|---|---|
| Resonance frequency | 6.8346875GHz |
| Loaded quality factor | >500 |
| Magnetic shield factor | 10000 |
| Cell dimensions | O.D=25mm,L=27mm |
| Buffer gas | $N_2$ at 9.5torr |
| Cavity temperature | $75^0C$ |
| Heating power (cavity) | 3.81W |
| Mag. Shield temperature | $65^0C$ |
| Heating power for Mag. Shield | 5.5W |
| Thermal constant | 3600s |

**III RF PACKAGE**

The RF circuits form an important part of Rb atomic clock. These circuits are required to produce from a VCXO, generally a 10MHz, a resonant atomic transition frequency. The transition frequency is usually a fractional frequency. The corrected VCXO frequency is the clock output. Since both frequencies must be derived from the same source, some sort of frequency synthesis involving frequency division, multiplication and mixing is clearly required. A schematic diagram of the 6.834GHz synthesizer is shown Fig-2(a) and the photograph of the developed synthesizer is given in Fig-2(b).

Since the frequency synthesis process is more easily done at lower frequencies (up to few MHz), it becomes convenient to divide the RF circuits into two parts –a frequency multiplier and a synthesizer. The VCXO frequency is 10MHz.This frequency is multiplied to 360MHz (the multiplication factor are 3,3 and 4).A frequency of 5.3125MHz is synthesized from the same 10MHz oscillator and mixed with multiplier output at 360MHz to generate 360MHz±5.3125MHz by amplitude modulation technique.

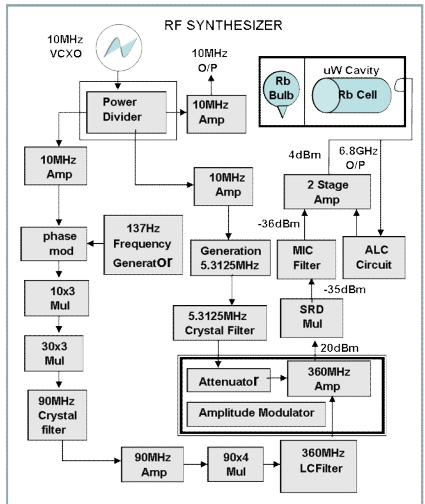
Fig 2 (a) - Schematic of 6.834GHz synthesizer.

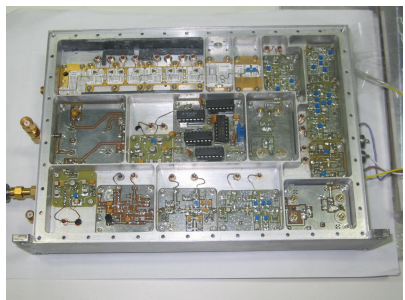
Fig 2 (b). 6.834 GHz RF Synthesizer

The amplifier used as amplitude modulator also serves as a pre-stage input amplifier for SRD multiplier which needs a minimum 20dBm input drive level. Due to nonlinear characteristic of SRD diode it generates the 19x360MHz-5.3125MHz= 6.834687500 GHz clock transition frequency. The 19th harmonic is selected by means of microwave band pass filter (Q) and is amplified from -34dBm to 4dBm by using two stage amplifiers CFY-67 HEMT. In fig-3 is shown the final RF spectrum of the synthesizer. The measured results of the RF systems are outlined in Table 2.

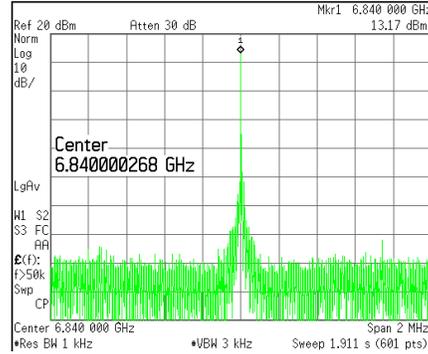
Fig 3- RF spectrum of synthesizer output

Table 2

Specifications of RF generation chain are as given

| Sr. No. | Parameter | Unit | Specification |
|---|---|---|---|
| 1 | O/P Frequency | GHz | 6.834687500 |
| 2 | O/P Power | dBm | +4 |
| 3 | O/P Power variation over temperature(-$10^0$-$60^0$C) | dBm | 0.01 |
| 4 | Harmonics | dBc | -30 |
| 5 | Spurious | dBc | -60 |
| 6 | SSB Phase Noise @ offset From center frequency<br>1Hz<br>10Hz<br>100Hz<br>1KHz<br>10KHz<br>100KHz | dBc | -28<br>-43<br>-68<br>-78<br>-88<br>-88 |

## IV SERVO SYSTEM

The servo system is used to lock the frequency of the VCXO to the atomic transitions. The input to the servo system is obtained from Rb physics package array of photo detectors. The output from the servo system is the error (correction) signal that goes to the control port of LO (VCXO) to correct the VCXO frequency. Before closing the loop the microwave frequency is varied through resonance, the error voltage developed at the output of the phase sensitive detector is shown in Fig 5.

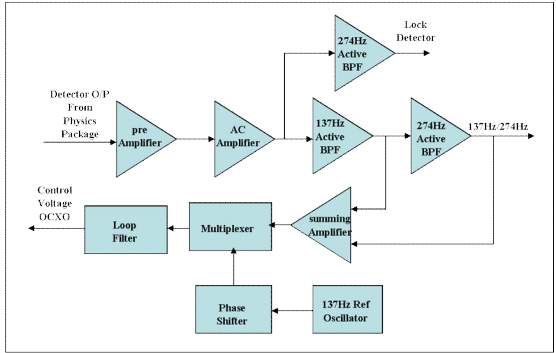

Fig-4 Functional block diagram of the servo section

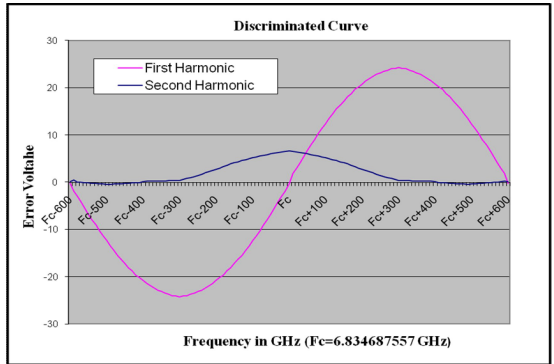

Fig 5- Error voltage at the output of the phase sensitive detector.

Table 3.
The measured results of Lock-In Amp. & Servo Section.

| Parameters | Unit | Measured results |
|---|---|---|
| DC Photocurrent | uA | 10 |
| Resonance line width | Hz | 600 |
| Gain | dB | 108 |
| Modulation depth | rad | 3.5 |
| Total Noise | pA/Hz | 30 |
| Min. Detectable Signal | dBm | -67 |
| Dynamic range | dB | 60 |
| Frequency Detection Sensitivity | V/Hz | 1/60 |
| Loop time constant | sec | 10s |

## V. PERFORMANCE

With the all subsystems running together, the frequency stability of the locked VCXO is $3\times10^{-11}/\sqrt{\tau}$ for $0<\tau<100s$ as shown in Fig-6. The entire system shown in Fig-1 has been tested for its short and long term frequency stabilities. The frequency of the stabilized 10MHz output is sampled every few seconds during a period of more than 24 hours. The short term-stability of the output signal has been measured at time constants ranging from 0.1s to 1000s. The results are shown in Fig-6. For time constant 0.1s to 200s, curve varies linearly with $\sqrt{\tau}$, attaining a value of $2\times10^{-11}$ at 1 sec. Minimum observed Allan deviation value is $8\times10^{-12}$ at 400 sec. and then it starts to increase. Fig-6 shows frequency stability plot of the free running OCXO and locked OCXO. The DVM Rb clock has frequency stability of $10^{-11}/s$ and the same phase noise specification as that of the crystal oscillator [8] (<-121dBc/Hz at 100Hz and <-141dBc/Hz at 1KHZ). Phase noise of the VCXO outputs at 10MHz in unlocked and locked conditions are shown in fig 7.

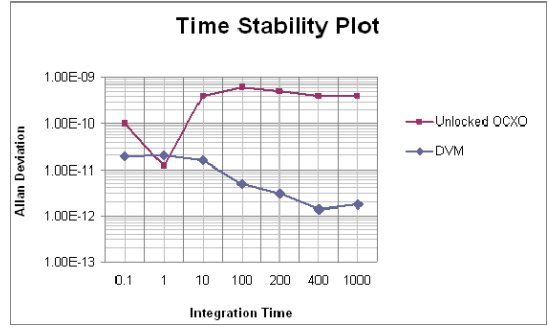

Fig 6- Allan deviation of unlocked and locked OCXO.

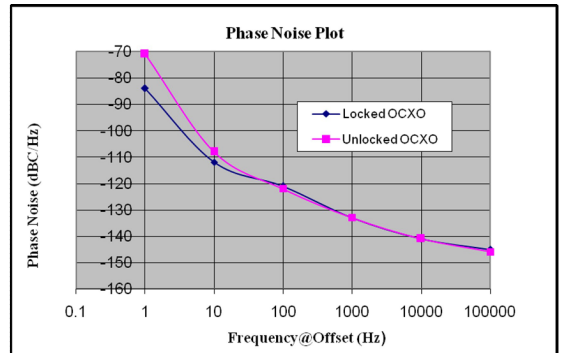

Fig -7 Phase noise plot of unlocked and locked VCXO.

## VI. CONCLUSION

We have discussed in this paper the work at NPL India and SAC for developing DVM of rubidium frequency standard. These devices are projected to have a volume about $406\times301\times160mm^3$, the power dissipation of 35watt and fractional frequency instability of the order of $1\times10^{-11}$ at one hour of integration. At present, the three critical subsystems, which make up the frequency reference standard, have been demonstrated. Together, they achieve a fractional frequency instability of $2\times10^{-11}/\sqrt{\tau}$ for $0<\tau<100s$, and improvements both in frequency stability performance and size reduction will be achieved as this work progresses. We have also explained the reasons of the mode changes in the Rb lamp. The CPT phenomenon has a part in the mode change. We have also explained that by placing the photo detector inside the microwave cavity the RF leakage is reduced.

**Acknowledgements:** We are thankful to Dr Kiran Kumar, SAC and Prof.R.C.Budhani, NPL India for their constant encouragement and guidance for the R&D work on Rb atomic clock.